\documentclass[namedreferences]{solarphysics}
\usepackage[applemac]{inputenc} 
\usepackage[hyperref,optionalrh,solaromanenum]{spr-sola-addons} 
\usepackage{graphicx}                    
\usepackage{color}                       
\usepackage{breakurl}                         
\def\fm{\hbox{$^{\rm m}$}}
\def\fs{\hbox{$^{\rm s}$}}

\begin{document}

\begin{article}

\begin{opening}

\title{Christian Horrebow's Sunspot Observations -- I. Life and published writings}

%
 \author[addressref={aff1}]{\inits{C.S.}\fnm{Carsten S\o{}nderskov } \lnm{J\o{}rgensen}}\orcid{https://orcid.org/0000-0002-0319-1207}
 \author[addressref={aff2,aff1},corref,email={karoff@phys.au.dk}]{\inits{C.}\fnm{Christoffer } \lnm{Karoff}}\orcid{https://orcid.org/0000-0003-2009-7965}
 \author[addressref={aff3}]{\inits{V.}\fnm{Valliappan Senthamizh } \lnm{Pavai}}\orcid{https://orcid.org/0000-0003-2413-7901}
 \author[addressref={aff3},email={rarlt@aip.de}]{\inits{R.}\fnm{Rainer }\lnm{Arlt}}

%

\address[id=aff1]{Stellar Astrophysics Centre, Department of Physics and Astronomy, Aarhus University, Ny Munkegade 120, 8000, Aarhus C, Denmark}
\address[id=aff2]{Department of Geoscience, Aarhus University, H{\o}egh-Guldbergs Gade 2, 8000, Aarhus C, Denmark}
\address[id=aff3]{Leibniz-Institut f\"ur Astrophysik Potsdam, An der Sternwarte 16, 14482 Potsdam, Germany}
\begin{abstract}
Between 1761 and 1776, Christian Horrebow made regular observations of sunspots from Rundet\aa{}rn in Copenhagen. Based on these observations he writes in 1775 that {\it it appears that after the course of a certain number of years, the appearance of the Sun repeats itself with respect to the number and size of the spots.} Thus, Horrebow hypothesized about the idea of a cyclic Sun several decades before Heinrich Schwabe discovered the solar cycle and estimated its period. This proves the ability of Horrebow as a sunspot observer. Here we present a general overview of the work of Christian Horrebow, including a brief biography and a complete bibliography. We also present a translation from Danish to English of his writings on sunspots in Dansk Historisk Almanak. These writings include tables of daily sunspot measurements of which we discuss the completeness.
\end{abstract}

%

\end{opening}

\section{Introduction}
Christian Pedersen Horrebow was born the parents of Anne Margrethe Rossings and Peder Nielsen Horrebow. His father was, at that time, the head of the Royal Danish Observatory at Rundet\aa{}rn, Copenhagen ($55^{\circ}40'53.00''\,$N, $12^{\circ}34'32.99''\,$E). When Peder Horrebow became too ill to properly manage his position in 1753, Christian Horrebow took over the daily management and was given the title of Professor of Mathematics and Natural Philosophy. After his father's death in 1764, he was officially appointed Professor of Astronomy, a position he held until his own death in 1776.

From 1761 to 1776, Christian Horrebow made regular position measurements of spots on the solar disk. These records constitute one of the most complete records of sunspots from the 18th century, together with the records by Staudacher and Flaugergues. Especially, cycle~2 is well covered by Horrebow's measurements \citep{1995SoPh..160..387H}.

Christian Horrebow's sunspot observations have been analysed by \citet{1859AN.....50..257T} who provided a catalogue of monthly mean values and by \citet{1873MiZur...4...77W} who used the values provided by Heinrich Louis d'Arrest, a professor at Copenhagen University at the time. Unfortunately, as noted by \citet{1995SoPh..160..387H} the two catalogues do not agree as \citet{1859AN.....50..257T} had a larger likelihood of calling something a sunspot group than d'Arrest. \citet{1995SoPh..160..387H} therefore made another catalogue of Christian Horrebow's sunspot records, this time stating the group sunspot number. This catalogue is still in use today. It is, however, unclear how \citet{1995SoPh..160..387H} estimated their group sunspot numbers. As we will describe in detail below, the notebooks contain both tables of measurements and illustrations of the sunspots. We will argue that most information is contained in the tables and that this information is then only illustrated in the sunspot drawings. Caution should therefore be taken when using the illustrations as reproductions of the Sun. 

The 11-year solar cycle is also known as the Schwabe cycle after Heinrich Schwabe who, based on 17~years of observations, suggested a 10-year periodicity in the appearance of groups of spotless days \citep{1844AN.....21..233S}. Schwabe writes: {\it From my earlier observations, which I have reported every year in this journal, it appears that there is a certain periodicity in the appearance of sunspots, and this theory seems more and more probable from the results of this year.}\footnote{Original: Schon aus meinen fr\"uheren Beobachtungen, die ich j\"ahrlich in dieser Zeitschrift mittheile, scheint sich eine gewisse Periodicit\"at der Sonnenflecken zu ergeben und diese Wahrscheinlichkeit gewinnt durch die diesj\"ahrigen noch an Sicherheit.} This can be compared to what Christian Horrebow wrote 69~years earlier in 1775 in Dansk Historisk Almanak, based on 14~years of observations (a translated version of this text can be found in the Appendix to this paper): {\it (...) thus it appears that after the course of a certain number of years, the appearance of the Sun repeats itself with respect to the number and size of the spots, and over time, it will also be worth noting whether the number and size of the spots on the Sun have an influence on the state of the air here on Earth.}\footnote{Original: (...) dog lader det til, at visse Aars Forl\o{}b kommer samme Solens Skikkelse igien i Henseende til M\ae{}ngden og St\o{}rrelsen af Pletterne, og ligeledes bliver det i Tiiden v\ae{}rdt at l\ae{}gge M\ae{}rke til, om ikke Pletternes Tal i Solen og St\o{}rrelsen haver nogen Indflydelse paa Luftens Tilstand hos os paa Jorden.} Christian Horrebow did, however, not estimate a period, though the 10-year periodicity appears clearly in his observations.

Lately, the sunspot observations by Christian Horrebow have received renewed attention due to the ongoing work on providing a proper calibration of the sunspot number records \citep{2014SSRv..186...35C, 2017SoPh..292....4S, 2019NatAs...3..205M}. The observations by Christian Horrebow play an important role in this calibration as Christian Horrebow's observations span a time with very few available observations \citep{1998SoPh..181..491H,2016SoPh..291.3061V}, except for the drawings by the German amateur astronomer Johann Caspar Staudacher \citep{2008SoPh..247..399A}. 

In this study, which is the first in a series of papers on the sunspot observations by Christian Horrebow, we make a reanalysis of Christian Horrebow's work on sunspots along with a biography, a bibliography and a presentation of Christian Horrebow's sunspot observations and the telescopes which he used for these observations. This allows us not only to provide an improved version of Christian Horrebow's sunspot records, but also to make a qualitative assessment of the quality of Christian Horrebow's work which can be used when comparing the sunspot records of Christian Horrebow to the sunspot records of Staudacher. In the Appendix we provide a translation of Christian Horrebow's writings in Danish about sunspots. We also provide a discussion section on the problems with Christian Horrebow's observations of the 1760 Venus transit and the number of sunspots on 23~October 1769. 

The second paper \citep[][hereafter Paper II]{paperII} in our series on the work on sunspot observations by Christian Horrebow will present a digitisation of the notebooks and new sunspot records covering 1761, and 1767 to 1777 based on these notebooks. These records will include sunspot positions and not just sunspot numbers as in the older records. This will allow us to construct a butterfly diagram based on the observations.

After the death of Christian Horrebow in 1776, the sunspot observations were only continued by his successor Thomas Bugge until the end of 1777. After that year, the only observations of sunspots found are during the 1787 solar eclipse and a few days in 1806--1807. These observations are, however, of very low quality.

\section{Biography}
Christian Horrebow was the fourth child out of~16. He graduated from high school (Danish {\it Gymnasium}) in 1732 and got a master's degree (Danish {\it Magister}) in 1738. In 1743, he was appointed {\it Professor Designatus} and started working at the observatory. In 1753, he was appointed Professor of Mathematics and Natural Philosophy and took over the daily management of the observatory, but it was not until his father died in 1764 that he was formally appointed Professor of Astronomy \citep{danskastronomy}

Christian Horrebow published most of his papers about the Sun in the journal {\it Dansk Historisk Almanak}. This journal dates back to the mid-16th century. A number of other publications where published in {\it Skrifter som udi det Ki\o{}benhavnske Selskab af L\ae{}rdoms og Videnskabers Elskere}, which was published by the Danish Academy of Sciences and Letters that had been formed in 1742. However, the only publication found on sunspots is from 1769 in {\it Skrifter som udi det Ki\o{}benhavnske Selskab af L\ae{}rdoms og Videnskabers Elskere} that is on sunspots.

In addition to Christian Horrebow's sunspot observations, it is also worth mentioning his work on the time measurements performed at the observatory. Until 1768, the clocks appear to have been of poor quality losing or gaining 15--20 seconds in one day. With new clocks in 1770, this accuracy was brought below one second a day. The times were corrected using stellar observations, as well as eclipses of the Sun and the moons of Jupiter.

\section{Bibliography}
For his time, Christian Horrebow was very productive in documenting and publishing his work. In this way he published, at least, 7~monographs in Latin and 28~articles in Danish (see list below). Additionally, 20~notebooks by Christian Horrebow containing mainly sunspot observations are preserved. The contents of these notebooks will be analysed in Paper~II. They are handwritten, with tables as in the articles and drawings for most cloudless days as well as comments about the observations in Latin. Whereas the notebooks cover 1761 and 1764--1776, the articles only cover 1768--1776. Observations from earlier years may have been dating as far back as 1740, but these are presumed to have been destroyed in 1807 during the English bombardment of Copenhagen. The observations of 1777 were not published, since they were not made under Christian Horrebow. All the sunspot observations in the articles can be found in the notebooks, but the notebooks additionally contain drawings in which we can identify further sunspots (see Paper~II). In turn, the articles provide comments on the observations that are only provided in Latin in the notebooks.

\subsection*{Monographs:} 
\begin{itemize}
\item {\it De parallaxi fixarum annua ex rectascensionibus, qvam post Roemerum et parentem ex propriis observationibus}, Havniæ, 1747
\item {\it Eccentricitatem Solis vel Terr\ae{} Constantem esse, eandemqve non singulis annis decrescere}, Havni\ae{}, 1749
\item {\it Dissertatio de Distantia Stellarum Fixarum}, Havni\ae{}, 1755
\item {\it De semita, quam in sole descripsit Venus per eundem transeundo die 6~junii ao. 1761}, Havni\ae{}, 1761
\item {\it Elementa Astronomi\ae{} Sph\ae{}ric\ae{} in Usum Pr\ae{}lectionum Conscripta}, Hafni\ae{}, 1762
\item {\it Specimen astronomi\ae{} practic\ae{}}, Hafni\ae{}, 1766
\item {\it Elementa Astronomi\ae{} Sph\ae{}ric\ae{} in Usum Pr\ae{}lectionum Conscripta}, Hafni\ae{}, 1783
\end{itemize}

\subsection*{Articles:}
 
(English translation of the title in brackets) 
\begin{itemize}
\item {\it Om Solens Eccentricitet} [On the Sun's eccentricity], Skrifter som udi det Ki\o{}benhavnske Selskab af L\ae{}rdoms og Videnskabers Elskere, S1 5.8, 1751
\item {\it Om Fixstjernernes Afstand Fra Jorden} [On the distance to the fixed stars], Skrifter som udi det Ki\o{}benhavnske Selskab af L\ae{}rdoms og Videnskabers Elskere, S1 6.6, 1754
\item {\it Om Atmosf\ae{}rens Hoyde} [On the height of the atmosphere], Skrifter som udi det Ki\o{}benhavnske Selskab af L\ae{}rdoms og Videnskabers Elskere, S1 7.3, 1758
\item {\it Om Form\o{}rkelser} [On transits], Dansk Historisk Almanak, 1761 
\item {\it Beretning om Jordski\ae{}lvet, som skeede d. 22 Dec. Ao. 1759} [Report on the earthquake that took place on 22~December 1759], Skrifter som udi det Ki\o{}benhavnske Selskab af L\ae{}rdoms og Videnskabers Elskere, S1 9.5, 1765
\item {\it Tidens Bestemmelse i Henseende til de Observationer, som skeede i Solen og Venere, da Venus Anno 1761. den 6te Junii passerede igiennem Solen} [Time corrections related to the observations of the Sun and Venus from the Venus transit of 6~June 1769], Skrifter som udi det Ki\o{}benhavnske Selskab af L\ae{}rdoms og Videnskabers Elskere, S1 9.6, 1765
\item {\it Observation af Solens Form\o{}rkelse, som indfaldt den 1ste Aprilis 1764. Giordt paa det Kongelige Observatorio i N\ae{}rv\ae{}relse af Hans Kongelige Majest\ae{}ts H\o{}ye Geheime-Conseil} [Observations of the Solar eclipse on 1~April 1764. His Royal Majesty's High Geheime-Conseil was present at the observations] Skrifter som udi det Ki\o{}benhavnske Selskab af L\ae{}rdoms og Videnskabers Elskere, S1 9.7, 1765 
\item {\it Christian Horrebows Reflexioner anlangende Veneris Drabant} [Reflexions by Christian Horrebow regarding a moon of Venus], Skrifter som udi det Ki\o{}benhavnske Selskab af L\ae{}rdoms og Videnskabers Elskere, S1 9.9, 1765 
\item {\it Om kalenderen} [On the calendar], Dansk Historisk Almanak, 1767
\item {\it Om den Julianske Calenderstiil} [On the Julian calender], Dansk Historisk Almanak, 1768
\item {\it Tabel II -- om deklinationer} [Table II -- on declinations], Dansk Historisk Almanak, 1768
\item {\it Om Soel-Pletter} [On sunspots], Skrifter som udi det Ki\o{}benhavnske Selskab af L\ae{}rdoms og Videnskabers Elskere, S1 10.20, 1769$^\ast$
\item {\it Tabel I -- om Solens distance} [Table I -- on the distance to the Sun], Dansk Historisk Almanak, 1769
\item {\it Tabel II -- Det sande d\o{}gn er l\ae{}ngere eller kortere end Middeld\o{}gnet} [Table II -- The true Day is longer or shorter than the mean day] Dansk Historisk Almanak, 1769
\item {\it Om Veneris Gang Igiennem Soelen} [On the Venus transit], Dansk Historisk Almanak, 1769
\item {\it Om Soelpletterne} [On sunspots], Dansk Historisk Almanak, 1770 (Note that the text in this paper is a copy of the article in {\it Skrifter som udi det Ki\o{}benhavnske Selskab af L\ae{}rdoms og Videnskabers Elskere} from 1769, but the observations are from 1768)$^\ast$
\item {\it Observationerne giordte Anno 1769 af Soelpletterne} [Observation of sunspots in 1769], Dansk Historisk Almanak, 1771$^\ast$
\item {\it Tabel I -- om hvordan 24 timer i stjerne- eller soltid omregnes til grader} [Table I -- on how 24~hours of sidereal time transfers into degrees], Dansk Historisk Almanak, 1771
\item{\it Tabel II -- om brugen af tabel I} [Table II - on the use of Table~I], Dansk Historisk Almanak, 1771 
\item {\it Observationerne giordte Anno 1770 af Soelpletterne} [Observation of sunspots in 1770], Dansk Historisk Almanak, 1772$^\ast$
\item {\it Om den forbedrede Stiil} [On the Gregorian calendar], Dansk Historisk Almanak, 1772
\item {\it Tabel I og II -- Om Solens op- og nedgange} [Table~I and~II -- On the rising and setting of the Sun], Dansk Historisk Almanak, 1772
\item {\it Observationerne giordte Anno 1771 af Soelpletterne} [Observation of sunspots in 1771], Dansk Historisk Almanak, 1773$^\ast$
\item {\it Observationerne giordte Anno 1772 af Soelpletterne} [Observation of sunspots in 1772], Dansk Historisk Almanak, 1774$^\ast$
\item {\it Om Soelpletterne i Aaret 1773} [On the sunspots of 1773], Dansk Historisk Almanak, 1775$^\ast$
\item {\it Om Soelpletterne i Aaret 1774} [On the sunspots of 1774], Dansk Historisk Almanak, 1776$^\ast$
\item {\it Om Soelpletterne i Aaret 1775} [On the sunspots of 1775], Dansk Historisk Almanak, 1777$^\ast$
\item {\it Om Soelpletterne i Aaret 1776} [On the sunspots of 1776], Dansk Historisk Almanak, 1778$^\ast$
\end{itemize}

Translations of the papers marked with asterisks are provided in the Appendix. Please note that in general the articles do not contain page numbers. The numbers given for {\it Skrifter som udi det Ki\o{}benhavnske Selskab af L\ae{}rdoms og Videnskabers Elskere} refer to the location in the library of the Danish Academy of Sciences and Letters. All monographs and articles are available from the Royal Danish Library. The notebooks that we will analyse in Paper~II are not listed in the catalogue of the Royal Danish Library, but are available from the Library of Mathematics, Aarhus University (which is part of the Danish Royal Library).

In the early 17th century, a number of almanacs were published in Denmark, mainly containing astrological predictions of wars and natural disasters. This was changed in 1685 when Ole R\o{}mer reformed the publication of almanacs in Denmark so that only Copenhagen University was allowed to publish an almanac. This {\it Dansk Historisk Almanak} included no astrology-related papers. Instead, the focus was on astronomy and geological surveys with papers listing things like times of sunrise and sunset, lunar phases and exact locations of various Danish cities. The front pages list the names of a number of professors and students at Copenhagen University, who are likely to have been responsible for the editing process. 

{\it Skrifter som udi det Ki\o{}benhavnske Selskab af L\ae{}rdoms og Videnskabers Elskere} was published by 
 Danish Academy of Sciences and Letters from 1745 to 1779 and included scientific papers covering both natural sciences and humanities by people like Pontoppidan (realist writer), Holberg (play writer) and Spidberg (geologist). It did not, however, include articles that were mainly data papers such as some of Christian Horrebow's papers in {\it Dansk Historisk Almanak}.
 
It general, the monographs are longer dissertations analysing a scientific problem, such as the distance to the stars. The articles in {\it Skrifter som udi det Ki\o{}benhavnske Selskab af L\ae{}rdoms og Videnskabers Elskere} are scientific papers, discuss e.g. the nature of the sunspots, while most of the articles in {\it Dansk Historisk Almanak} are data papers, mainly presenting the sunspot observations.

Christian Horrebow's articles about sunspots include 9~articles in {\it Dansk Historisk Almanak} (1761, 1770--1778), though the one from 1771 is only a reference to an article in {\it Skrifter som udi det Ki\o{}benhavnske Selskab af L\ae{}rdoms og Videnskabers Elskere}, which again has an introductory text identical to the article in {\it Dansk Historisk Almanak} from 1772. The articles in {\it Dansk Historisk Almanak} include tables with sunspot observations from two years before the publication, so the 1770 {\it Dansk Historisk Almanak} contains observations from 1768, the 1772 publication contains observations from 1770, and so on. The article from 1761 is on observations of the Venus transit of 6~June 1761, but is also relevant for the sunspot observations. 
 
Even though Christian Horrebow is the author of all the monographs, articles and notebooks, it is clear that he neither did all the work nor all the writing. For example, all observations in the notebooks are associated with an observer. Only in a limited number of cases is Christian Horrebow the observer (see Paper~II for details). It is also clear that Christian Horrebow was not the author of the 1774 article in {\it Dansk Historisk Almanak}, which was written by R.~Jansen at the request of Christian Horrebow. In general, no author is indicated for the articles in {\it Dansk Historisk Almanak}. In the article in {\it Skrifter som udi det Ki\o{}benhavnske Selskab af L\ae{}rdoms og Videnskabers Elskere}, where Christian Horrebow is indicated as the author. He writes that he will continue to publish his sunspot observations in {\it Dansk Historisk Almanak}. It is therefore reasonable to assume that Christian Horrebow was indeed the author of the articles in {\it Dansk Historisk Almanak}, except for 1774, 1777 and 1778 (since he died in 1776). Here we need to remembered that the articles contain observations from two years earlier, so Christian Horrebow did take part in the observations.

\section{Telescopes}
In the 1770 {\it Dansk Historisk Almanak} article, Christian Horrebow writes: ``I have a large volume of observations of sunspots performed over 30~years, in part by myself, in part by my assistants in these observations.'' These observations, however, may have been lost during the 1807 bombardment of Copenhagen. It therefore appears that Christian Horrebow had been observing sunspots for quite some time before the year 1761, for which we have the first surviving protocol. The observations can be divided into two periods: one going from 1761 to 1766 and the other from 1767 to 1776 (possibly including 1777 where Christian Horrebow's method was continued). 

During the first period, the observers always used a quadrant ({\it Qvadrans}, in the notes using the ablative {\it Qvadrante}, which means {\it by means of a quadrant}) to measure the vertical distance of the sunspots to the Sun's upper or lower limb. To obtain the horizontal distance, they either used {\it Qvadrans} or {\it Rota Meridiana} (meridian circle). The latter could only be used at noon, since it was positioned to observe in the meridian of the observatory. A third instrument called {\it Machina Parallactica} (parallactic) was also used for three weeks in May/June of 1761 to find the distance of the sunspots to the lower limb. Except for these observations, the instrument appears in connection with sunspot observations in the notebooks only for two more days during the second period, namely 2 and 3 June 1769.

During the second period from 1767, an instrument called {\it Machina \AE{}quatorea} (equatorial) was the primary instrument used for the observations of sunspots. It was also in 1767 that the Gregorian telescope was inaugurated for making detailed drawings of sunspots.

We do not have much information about the quadrant, except that different versions appear in the notebooks: {\it Qvadrans/Quadrans}, {\it Qvadrans Paris} and {\it Qvadrans London}. \citet{bugge} also describes a 6-foot and a 3-foot quadrant ({\it Quadrans muralis} and {\it Quadrans mobile}), but it is unclear, if these refer to any of the quadrants in the notebooks.
{\it Rota Meridiana}, {\it Machina Parallactica} and {\it Machina \AE{}quatorea} had all been developed by Ole R\o{}mer and described in Latin in Basis Astronomi\ae{} \citep{peder}. It is likely that all three telescopes were lost in the great fire of 1728 in Copenhagen and then later rebuilt.

It should be noted that some drawings from 1761 have the vertical axis pointing down, while for all other years it is pointing up. As mentioned earlier, we do not have much information about the quadrant, but for {\it Rota Meridiana}, there are several drawings of views seen through the telescope. This was presumably a yearly calibration of the instrument, since they were performed at about the same time each year. The instrument had a grid with 7~vertical lines and 3~horizontal lines as can be seen in Figure~1, and all lines seem fixed. There are, however, other observations, e.g. on 17~May 1761, which give the impression that the horizontal cross hairs are movable. As mentioned earlier and indicated by the name of the instrument, {\it Rota Meridiana} could only observe in the meridian of the observatory. Its use for sunspot observations was therefore limited by the requirement for good weather exactly at noon. 

Very little has been found about how {\it Machina Parallactica} worked, only that in 1771 (starting on 2~May), it appears to have had a vertical line in the field of view when used to observe a comet (Notebook~9, see Paper~II).

\begin{figure} 
\centerline{\includegraphics[width=0.8\textwidth,clip=]{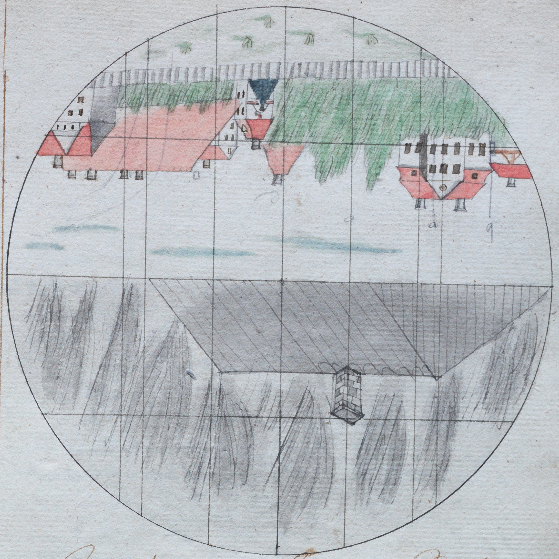}}
\caption{Example of a drawing based on observations with {\it Rota Meridiana}.}
\end{figure}

\subsection{Machina \AE{}quatorea}
During 1767, the staff at the observatory started using a new instrument called {\it Machina \AE{}quatorea}, which was developed by Ole R\o{}mer, as was the case for {\it Rota Meridiana}. A possible reason why {\it Machina \AE{}quatorea} had not been in use earlier is that it had to be rebuilt after the great fire of 1728 in Copenhagen, which also reached the observatory. After the fire, Peder Horrebow had a hard time getting the observatory up and running again with limited funding.

With {\it Machina \AE{}quatorea}, Christian Horrebow and his assistants could observe sunspots at any time of day and the two sets of coordinates could be obtained right after each other. It appears that they used much of 1767 to make corrections to the positioning of this instrument, as there are several descriptions of this in Danish in the protocol for that year. The last correction was made on 25~September 1767.

For the remainder of Christian Horrebow's time as  head of the observatory, {\it Machina \AE{}quatorea} was the only instrument used to obtain coordinates for the observed sunspots. After Horrebow's death in 1776, Rasmus Leivog (under Thomas Bugge's leadership) continued to make observations with the method developed by Christian Horrebow until the end of 1777. From July 1775 and onwards, Rasmus Leivog was the primary observer, so the last observations attributed to Horrebow were actually performed by Rasmus Leivog (nearly all the observations in the protocols were made by assistants -- Christian Horrebow is only mentioned in the notebooks for about 50 of the approximately 1800 sunspot observations we have).

\begin{figure}
\centerline{\includegraphics[width=5.5cm,clip=]{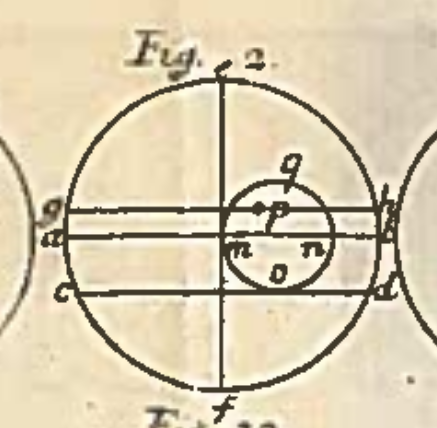}}
\caption{Example of how sunspot observations are undertaken with {\it Machina \AE{}quatora}.}
\end{figure}

A detailed description of the method developed by Christian Horrebow for sunspot observations with {\it Machina \AE{}quatorea} can be found in the 1770 {\it Dansk Historisk Almanak} article. The field of view contains a grid with 3 vertical lines and 1 horizontal line, see Figure~2. The line `gh' could be moved up and down, while the other lines were fixed. This allowed the observer to first record all horizontal coordinates for the sunspots simply by letting time pass and write down when the sunspot made contact with the vertical line `ef'. After this, the $y$-coordinates could be determined by moving `gh' to the sunspot and recording the distance to the line `cd' placed at the lower limb of the Sun. The image seen with this instrument is rotated by 180~degrees, so we have the Sun moving in the correct direction. Christian Horrebow does not mention this explicitly, but his naming of the `preceding', `lower' and `upper' limbs of the Sun fits this orientation. 

In general, we have very little information of the physical dimensions of the telescopes, such as aperture sizes and focal lengths. Such information can be found for earlier instruments constructed by Ole R\o{}mer in \citet{peder} and for later instruments in Thomas Bugge's notes on instruments available at the observatory before the bombardment of Copenhagen in 1807 \citep[reproduced in ][]{danskastronomy}.

\subsection{The Gregorian Telescope}
While {\it Machina \AE{}quatorea} provided detailed locations of sunspots on the solar disk, the drawings accompanying these observations are often just sketches. To make detailed drawings of sunspots, Christian Horrebow and his assistants used a Gregorian Telescope of 2 feet (0.63 m) in length. The Gregorian Telescope was first used at the Venus passage and solar eclipse in June 1769. The use of the Gregorian Telescope subsequently became more regular, and from 1770, is was routinely used. The drawings made with the Gregorian Telescope are rotated by 180~degrees compared to the drawings made with {\it Machina \AE{}quatorea}. This is illustrated in Figures~2 and~3 with drawings from 10~November 1770, made with the two instruments.

 \begin{figure}
 \centering
 \begin{minipage}[c] {6cm}
    a    \includegraphics[width=5.5cm]{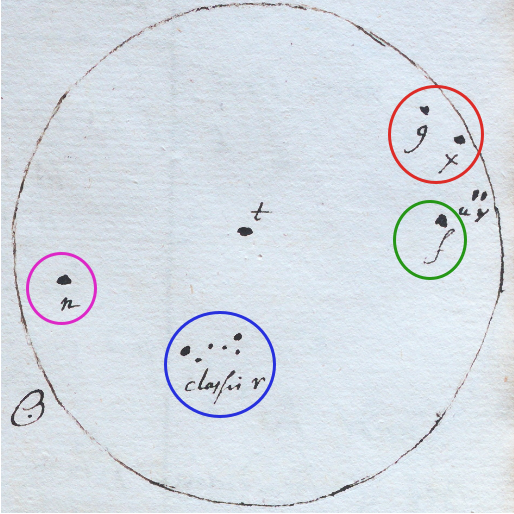} 
 \end{minipage}
  \begin{minipage}[c] {6cm}
     b    \includegraphics[width=5.5cm]{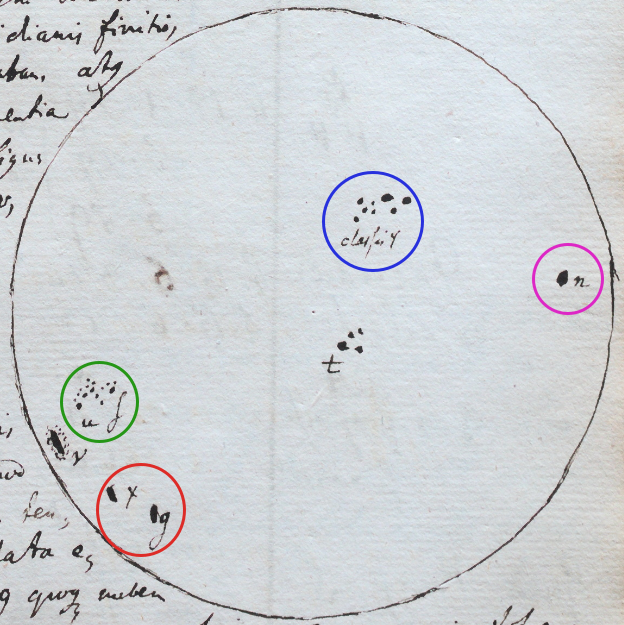} 
 \end{minipage}
 \caption{Illustration of the relative orientation between {\it Machina \AE{}quatora} (panel a) and the Gregorian Telescope (panel b). It is clear that (b) is rotated by 180~degrees with respect to (a).}
 \end{figure}
 
\section{Discussion}
In the 1776 article in {\it Dansk Historisk Almanak}, Christian Horrebow noted that ``They [the sunspots] have not been scarce since that time [1713], but in no year has there been such an abundance of them as in 1769, where a large number of them were regularly seen on the Sun, in particular on October~23, on which day I was able to count over 60~spots on the Sun, as otherwise the highest number anyone has seen at once only amounts to~50. Because the Sun's appearance was so unusually altered on that day on account of the numerous spots, it has been included on the chart of spots.'' The figure is reproduced in Figure~4 and one can indeed count 60~spots on 23~October 1769. 
\begin{figure} 
	\centerline{\includegraphics[width=1.0\textwidth,clip=]{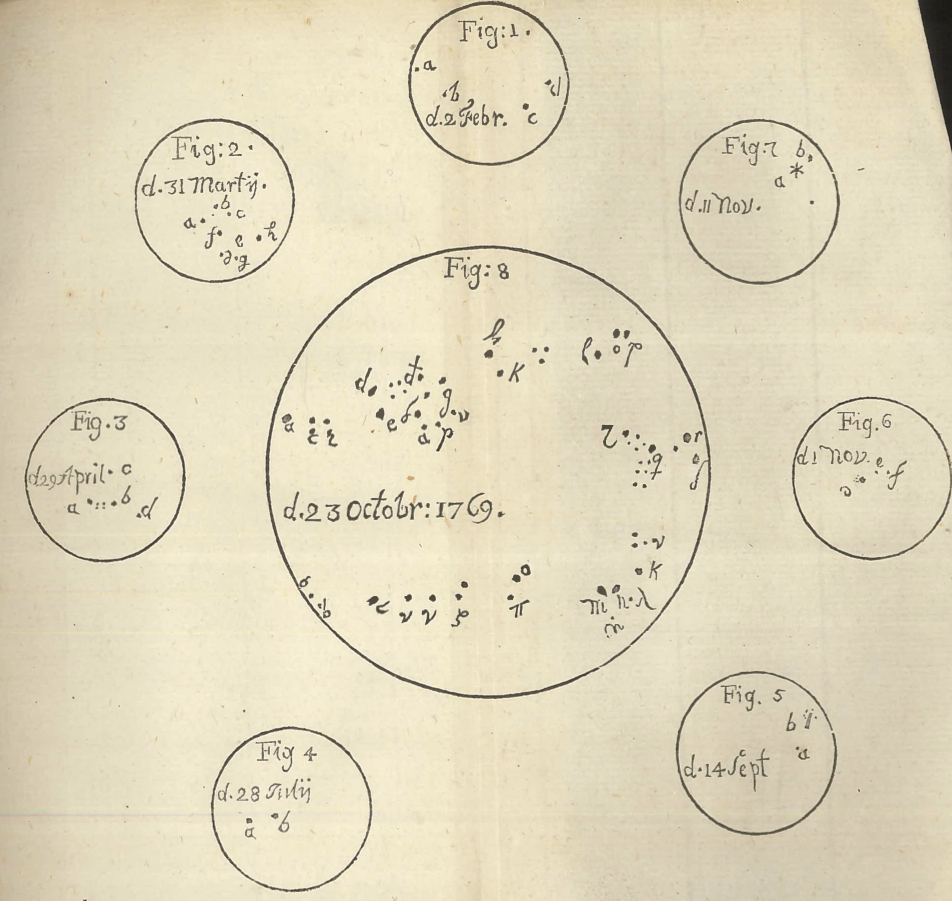}}
	\caption{Reproduction of the figure in the 1776 article in {\it Dansk Historisk Almanak}. The drawing in the middle is from 23~October 1769, which is the day when Christian Horrebow saw the largest number of spots on the Sun.}
\end{figure}
\begin{figure} 
	\centerline{\includegraphics[width=1.0\textwidth,clip=]{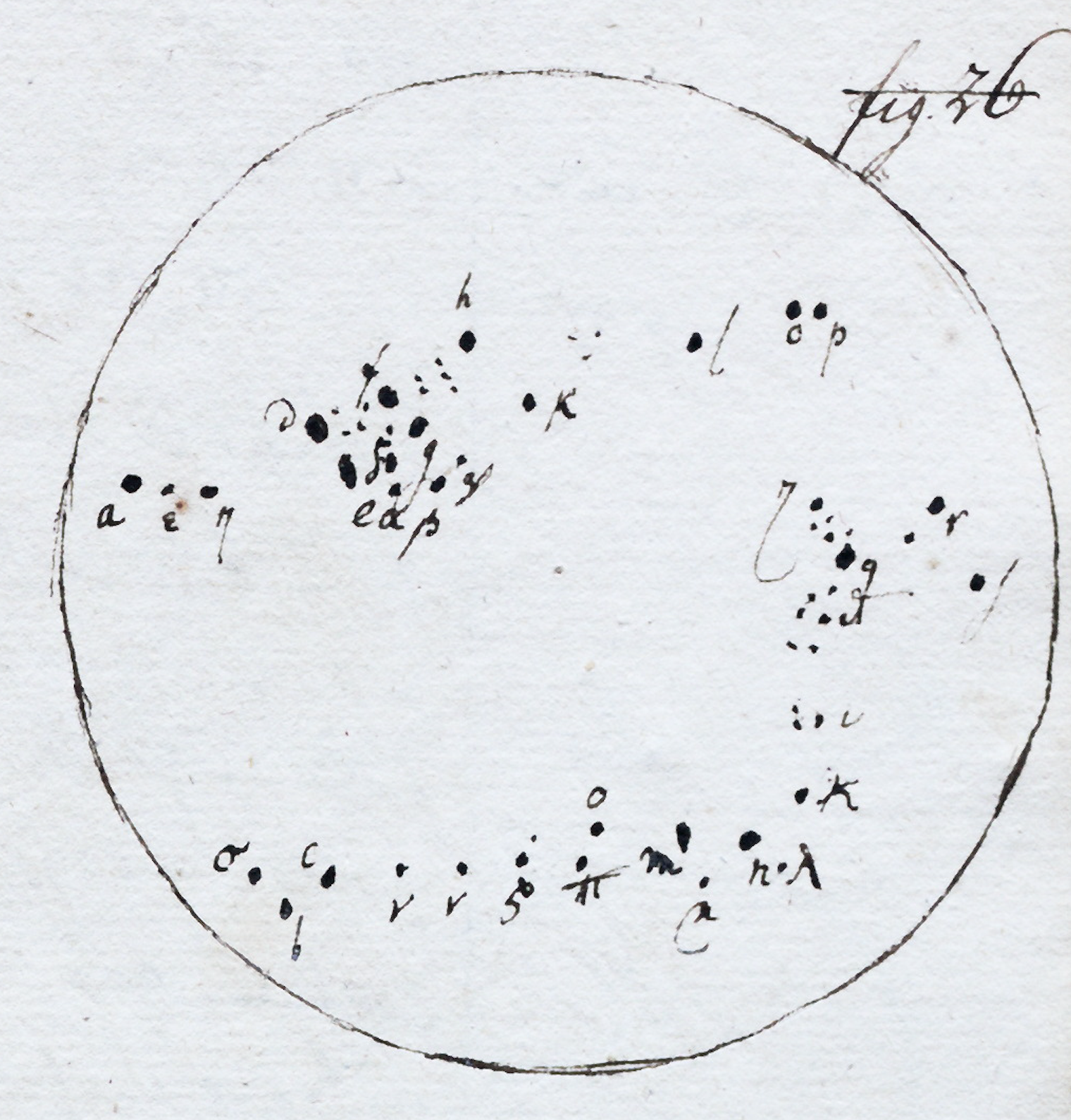}}
	\caption{The drawing from 23~October 1769, but this time reproduced from the notebook. See Paper~II for a detailed description of the notebooks.}
\end{figure}

The 23~October 1769 drawing shows a similar number of sunspots as seen on the most sunspot rich days in the 20th century. In general, the mid-18th century is the period with the largest disagreement between different calibrations \citep[see i.e.][]{2014SSRv..186...35C, 2016SoPh..291.2653S, 2016SoPh..291.2685U, 2016SoPh..291.3061V}. A proper cross-calibration between the new Horrebow sunspot record, published in Paper~II, and other sunspot number records however, has to be performed before any conclusion can be drawn.

We should add that Staudacher only observed around 25 spots on 25~October (no observations were made on the 23rd). He did, in fact, observe more spots, around~30, during the preceding Carrington rotation on 26~September \citep{2008SoPh..247..399A}.

We think that one reason why Christian Horrebow's idea about a cyclic Sun has not attracted more attention is likely to be him missing a reputation as a talented observer of his time, due to his observations of the Venus transit in 1761. At the time, observations were conducted with {\it Machina \AE{}quatorea} at the observatory after sunrise, when the transit had already begun, until the end of the transit a few hours later. Unfortunately, no corrections were applied to the transit times before they were sent to J\'er\^ome Lalande in Paris. This resulted in the observations being off by 2\fm51\fs. The problem was described in the 1765 article in {\it Skrifter som udi det Ki\o{}benhavnske Selskab af L\ae{}rdoms og Videnskabers Elskere}, but the article came far too late and Lalande discarded the observations. The story earned Christian Horrebow a reputation as being incompetent \citep{danskastronomy}. This is also reflected by the fact that Christian Horrebow was not selected by the Danish government to lead the 1769 Norway expedition for observing the next Venus transit \citep{danskastronomy}. Also, Christian Horrebow published his work in a little-known Danish journal.

The fact that Christian Horrebow suggested that the Sun repeats itself with respect to the number and size of the spots, 69~years before Heinrich Schwabe discovered the solar cycle and estimated its period, indicates that Horrebow was a more talented observer than he was given credit for by his contemporaries.

Due to the blackletter typesetting and to the fact that the Danish language has changed considerably since the mid-18th century, the articles in {\it Dansk Historisk Almanak} have been translated using a two-step procedure, where the articles were first translated into modern-day Danish and then into English. A few sentences and one section are in Latin and have also been translated.

Obvious errors have been removed from the tables, but only when there were clear typesetting errors. Mistakes that could be misunderstandings have all been kept.

\subsection{Summary}
During his time as director of Rundet\aa{}rn, Christian Horrebow made systematic sunspot observations. Regular observations span the time from 1761 to 1776. {\it Machina \AE{}quatorea} was inaugurated in 1767 and used as the primary instrument for the remaining sunspot observations overseen by Christian Horrebow. These measurements consist mainly of $x$ and $y$ coordinates of the sunspots, accompanied by drawings. 

We argue that the coordinates in the tables are of a higher quality and contain more information than the accompanying drawings. This also means that we can use the coordinates to construct a new sunspot record, which includes positions based on Christian Horrebow's observations, even for days when only a table and no drawings are available.

In general, we argue that Christian Horrebow was more productive than previously given credit for, with detailed observations being performed over more than a decade under his leadership. The fact that the sunspot observations were abandoned only one year after his death indicates that he was the driving force behind them.

Christian Horrebow suggested in 1775 that the Sun repeats itself with respect to the number and size of spots after a certain number of years, i.e.\ that the Sun is cyclic. We speculate that the reason why Christian Horrebow's findings have not received more attention is to be found in that Christian Horrebow published his findings in Danish and in the tarnished reputation he had earned due to an unfortunate matter with the 1761 Venus transit. The last reason is ironic as Christian Horrwbow's idea about a cyclic Sun exactly highlight him as a good observer.

\begin{acks}
We would like to thank the referee for thoughtful comments which significantly improved the paper. The translation of Christian Horrebow's published papers on sunspots was done by Lenore Messick, Events and Communication Support, Aarhus University. The project has been supported by the Villum Foundation. Funding for the Stellar Astrophysics Centre is provided by the Danish National Research Foundation (grant agreement no.\ DNRF106), funding for VSP is provided by the Deutsche Forschungsgemeinschaft (grant no.\ AR355/12-1). We are very grateful to Susanne Elisabeth N\o{}rskov, who has been most helpful throughout the project with finding the notebooks and other historical texts. We are also grateful to Kristian Hvidtfelt Nielsen, who helped with interpreting the historical texts.
\end{acks}

%
%
%

\end{article} 
\end{document}